\documentclass[twocolumn,showpacs,preprintnumbers,amsmath,amssymb]{revtex4}

\usepackage{graphicx}
\usepackage{dcolumn}
\usepackage{bm}

\def\D{\partial}
\def\grad{\nabla}
\def\div{\nabla \cdot}
\def\rot{\nabla \times}

\def\inv{^{-1}}

\def\const{\mbox{const.}}

\def\Av.#1{\overline{#1}}
\def\Eq.#1{(\ref{eq:#1})}
\def\Fig.#1{Fig.\ref{fig:#1}}   

\def\eq{\begin{eqnarray}}
\def\qe{\end{eqnarray}}

\def\eqnn{\begin{eqnarray*}}
\def\qenn{\end{eqnarray*}}

\def\nn{\nonumber}

\def\bem{\bm{m}}
\def\bn{\bm{n}}

\def\br{\bm{r}}

\def\simge{\;\lower3pt\hbox{$\stackrel{\textstyle >}{\sim}$}\;}
\def\simle{\;\lower3pt\hbox{$\stackrel{\textstyle <}{\sim}$}\;}

\def\->{\rightarrow}
\def\=>{\Rightarrow}
\def\<->{\leftrightarrow}
\def\<<{\ll}
\def\>>{\gg}

\def\bm#1{\mbox{\boldmath $#1$}}

\def\lrM#1{\left\{#1\right\}}
\def\lrS#1{\left(#1\right)}
\def\lrF#1{\left|#1\right|}
\def\lrA#1{\left\langle #1 \right\rangle}

\def\mycomment#1{}

\def\f#1#2{\frac{#1}{#2}}
\def\der#1#2{\f{\D #1}{\D #2}}
\def\fder#1#2{\f{\delta #1}{\delta #2}}

\begin{document}

\title{Numerical simulation of the twist-grain-boundary phase of chiral liquid crystals}
\author{Hiroto Ogawa}\email{hiroto@cmpt.phys.tohoku.ac.jp}
\author{Nariya Uchida}
\affiliation{
Department of Physics, Tohoku University, Sendai, 980-8578, Japan
}

\date{March 8, 2006; published June 19, 2006 [Phys. Rev. E {\bf 73}, 060701 (R)(2006)]}

\begin{abstract}
We study the structure of the twist-grain-boundary phase of 
chiral liquid crystals
by numerically minimizing the Landau-de Gennes free energy.
We analyze the morphology of layers at the grain boundary, 
to better understand the mechanism of frustration between
the smectic layer order and chirality.
As the chirality increases, the layer compression energy 
strongly increases while the effective layer bending rigidity is 
reduced due to unlocking of the layer orientation and the director.
This results in large deviation of the layer morphology from that of 
Scherk's first minimal surface and linear stack of screw dislocations.
\end{abstract}

\pacs{
61.30.Jf, 
61.72.Mm, 
61.20.Ja  
}
\maketitle

Frustration causes complex structural patterns 
in a variety of materials and over different lengthscales. 
To resolve the frustration,
the equilibrium patterns often contain topological defects.
Liquid crystals are some of the richest materials 
providing a variety of frustrated defect phases.
Defects are inevitably formed by frustration between 
the smectic order and chirality, because the periodic 
layer structure and continuous helical structure are 
geometrically incompatible. As a result, the frustration 
gives rise to a set of screw dislocations.
The simplest stable phase containing such defects 
is the twist-grain-boundary (TGB) phase, the existence of which was 
predicted using the analogy with the Abrikosov vortex lattice
of type-II superconductors~\cite{Renn}.
In this phase, smectic slabs of a certain length (grains) 
are twisted by a certain angle  from their neighbors
and separated from them by a narrow region (grain boundary)
in which screw dislocations are aligned.
\par
After its theoretical prediction~\cite{Renn} and 
experimental confirmation~\cite{Goodby},
study of the TGB phase was extended to other 
chiral frustrated phases and transitions between them.
Theoretically, thermal fluctuation changes the transitions 
between smectic-A (Sm-A), TGB$_{\rm A}$, TGB$_{\rm C}$ and cholesteric ($N^*$) phases 
from second-order to first-order.
In parallel to the vortex liquid phase of superconductors, fluctuation 
induces the chiral line ($N_{\rm L}^*$) phase with a melted defect lattice, 
which was theoretically predicted~\cite{NL} and 
experimentally confirmed~\cite{Chan,Navailles}.
Many other chiral frustrated phases have been 
found and reported~\cite{Kitzerow,Yamamoto}.
\par
On the other hand, the spatial structure of the TGB phase, especially that 
of a grain boundary, still has much to be understood.
Kamien and Lubensky~\cite{Kamien} showed that the grain 
boundary structure 
is well described by a linear stack of screw dislocations (LSD) 
if the twist angle $\alpha$ is close to zero.
For small $\alpha$, the layer structure is also approximated 
by Scherk's first minimal surface.
However, the minimal surface defined by vanishing mean curvature $H$ 
is achieved when the bending elasticity is the only contribution 
to the free energy. 
Deviation from the minimal surface should occur due to the other 
elastic effects, namely the layer compression and the twist Frank
energies.
Also, spatial variation of the density near the TGB defect core is 
not considered in the previous analytic studies, and
the interplay between these effects remains to be 
investigated~\cite{Linear,NonLinear,Santangelo}.
Study of the TGB structure for large twist angles should be 
also helpful for understanding the structure of more complex phases, 
such as the $N_{\rm L}^*$ and smectic blue phases~\cite{Kitzerow,DiDonna}.
\par
It might be worth mentioning that the TGB structure is also exhibited by
other materials such as twisted lamellae of block copolymer melts~\cite{Duque}
and defect-containing Turing patterns in a reaction-diffusion system~\cite{ADEWIT}.
The mechanism that determines the defect structure depends on 
the material, and the uniqueness of the liquid crystal TGB structure 
is not yet clarified.
\par
In this Rapid Communication, we study the core structure of a grain boundary
by numerical minimization of the Landau-de Gennes free energy.
Thus, compared to previous molecular simulations
of the TGB phase~\cite{Memmer,Allen}, 
our results allow more direct comparison with analytical results. 
Also, the previous simulations suffer from severe finite-size effect 
because the size of the simulation box and the grain size $\ell_b$
are generally incommensurate with each other.
By focusing on a single grain boundary, we become free from this problem
and can study the core structure with a higher resolution. 
The neglected grain boundary interaction would not affect the core
structure except in the vicinity of the TGB-$N^*$ transition.
The difference between the simulation result
and the model surfaces (i.e., Scherk's first surface and the LSD)
is analyzed as a function of the twist angle $\alpha$.
We discuss the reason for the deviation in terms of the 
coupling between the smectic order parameter and the director.
\par
The free energy we utilize is
the covariant Landau-de Gennes model expressed as~\cite{Renn,deGennes}
\eq
F&=&F_{D.W.}+F_{int}+F_{Frank},\\
\label{eq:L-deG}
F_{D.W.}&=&\int d\br \f{g}{4} \lrS{\f{\tau}{g} + |\Psi|^{2}}^{2},
\\
F_{int}&=&\int d\br \f{B}{2}\lrF{{\lrS{\grad-iq_{0}\bn}\Psi}}^{2},
\qe
\eq
F_{Frank}&=&
\int d\br \left\{ 
\f{K_{1}}{2}(\div\bn)^{2} \right. 
+\f{K_{2}}{2}\lrS{\bn\cdot\rot\bn-k_{0}}^{2}
\nn\\
&&
+\left. \f{K_{3}}{2}(\bn\times\rot\bn)^{2} 
\right\},
\qe
where $\Psi$ is the smectic (complex) order parameter
combining the density modulation ($\propto {\rm Re}\, \Psi$) 
and layer displacement, and $\bn$ is the director.
The double-well potential $F_{D.W.}$ 
with the dimensionless temperature $\tau$ 
controls the order-disorder (TGB--$N^*$ or Sm-A--$N^*$) transition. 
The interaction part $F_{int}$ with the coupling constant $B$
fixes the layer thickness $d$ to $2\pi/q_{0}$.
The third term is the Frank elastic energy.
Although the ratios between the Frank elastic constants $K_i$
affect the dislocation core structure~\cite{Kralj},
we set all the Frank constants to the same value $K$
and focus on the role of chirality and 
coupling between $\Psi$ and $\bn$.
The correlation length of the order parameter $\Psi$ is defined as
$\xi=\sqrt{B/|\tau|}$, the director penetration 
length $\lambda=\sqrt{Kg/B|\tau|}$ and the 
Ginzburg parameter $\kappa \equiv \lambda/\xi =\sqrt{gK}/q_{0}B$.
The TGB phase is stable only when $\kappa > 1/\sqrt{2}$ according 
to the mean-field theory~\cite{Renn}.
\begin{figure}
\includegraphics[width=70mm]{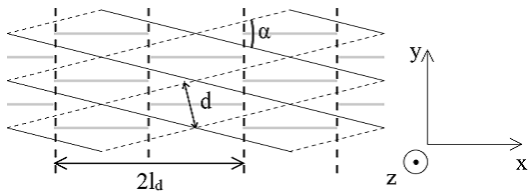}
\caption{Single twist grain boundary structure.
The broken and gray lines indicate the parallel screw dislocations 
and the smectic layers in the grain boundary plane ($z=0$).
The solid and dotted lines are the layers at $z \to \pm\infty$, respectively.
}
\label{fig:GB}
\end{figure}
\par
The numerical minimization is performed in 
a $L_{x} \times L_{y} \times L_{z}$ simulation box,
with the $y$- and $z$-axes identified with the direction 
of dislocations and the helical axis, respectively.
To fix the transverse dimensions $L_x$ and $L_y$,
we fix the twist angle per grain boundary $\alpha$ as 
an independent parameter, instead of the chirality $k_0$. 
Then, two smectic slabs sandwiching a single grain boundary 
satisfy a two-dimensional (2D) crystalline symmetry with the periods 
$\ell_x = d/\sin(\alpha/2)$ and $\ell_y = d/\cos(\alpha/2)$ (see Fig. 1).
Thus we can assume the periodic boundary condition in the 
transverse directions. We set $L_x=\ell_x$ and $L_y=\ell_y$
to save computation time.
The layer orientation changes by the angle $\alpha$
along the helical axis, which is imposed as a boundary 
condition at $z=0$ and $z=L_z$ as follows.
The director is set to $\bn= \bn_\pm = (\pm\sin(\alpha/2), \cos(\alpha/2), 0)$
at $z=L_z$ and $z=0$ respectively, while the smectic order parameter
at these boundaries are connected by inversion with respect to 
the plane $x=L_x/2$: $\Psi(x, y, 0)=\Psi(L_x-x, y, L_z)$.
The latter is compatible with the 2D crystalline symmetry
in the $xy$ plane.
For finite $L_z$,
these boundary conditions induce interaction of the grain boundary 
and its images. However, the interaction is expected to decay 
exponentially~\cite{Renn} and hence our boundary condition
gives a good approximation if $L_z/2 \ge \xi,\lambda$. 
We will use $L_z=2d$ in the simulation.
\par
The free energy is minimized by solving the time-dependent 
Ginzburg-Landau (TDGL) equations,
\eq
\der{\Psi}{t}&=& - \Gamma_{\Psi} \fder{F}{\Psi},
\label{eq:TDGLPsi}\\
\der{\bn}{t}&=&- \Gamma_n (\bm{1}-\bn\bn) \cdot \fder{F}{\bn},
\label{eq:TDGL2}
\qe
where $\Gamma_{\Psi}$ and $\Gamma_n$ are appropriate kinetic 
constants and the factor $\bm{1}-\bn\bn$ in \Eq.{TDGL2} 
ensures that $|\bn|^{2}=1$. 
For the initial condition we take two flat smectic slabs 
with sinusoidal order parameter profiles and layer normals
identical to $\bn_{\pm}$ for $z>L_z/2$ and $z<L_z/2$ (respectively).
Since we fixed $\alpha$ instead of $k_0$, the chirality is
determined by minimizing $F_{Frank}$ with respect to $k_0$ as
\eq
k_0=\f{1}{V}\int d\br\lrS{\bn\cdot\rot{\bn}},
\label{eq:updatek0}
\qe
which is calculated from the director configuration.
In the equilibrium, it is a function of the temperature $\tau$ 
and the twist angle $\alpha$: $k_0=f(\tau, \alpha)$.
Inverting this relation, we can find the equilibrium twist angle 
for given temperature and chirality as $\alpha=g(\tau,\; k_{0})$.
We use the parameter set $\tau=-0.02, g=1, B=0.2, K=0.02$ and $d=16$
with the unit mesh size $\Delta x=1$, unless otherwise stated.
For this choice, the correlation length and the penetration length 
are $\xi=0.44d$ and $\lambda=0.80d$ so that the conditions
$L_{z}/2 \ge \xi, \lambda$ and $\kappa > 1/\sqrt{2}$ are satisfied.
The kinetic coefficients are chosen as $\Gamma_\psi=\Gamma_n=0.1$ 
with the time step $\Delta t=1$.
Our criterion of equilibration is that
the free energy difference $F_i(t+t_0)-F_i(t)$ 
is lower than 0.01\% of the characteristic 
amount of $F_i$ for $\forall i$, where the index $i$
indicates each free energy component
and the relaxation time $t_{0}$ is determined by
the fitting $F(t) \sim \exp(-t/t_{0})+\const$ at
the early stage $t \le 2000$.
To prevent a trapping by local free energy minima, 
we also added a small noise which is gradually 
reduced to zero.
The equilibrated grain boundary structure for $\alpha=50$$^\circ$ 
is shown in \Fig.{SNAP}.
\begin{figure}
\includegraphics[width=75mm]{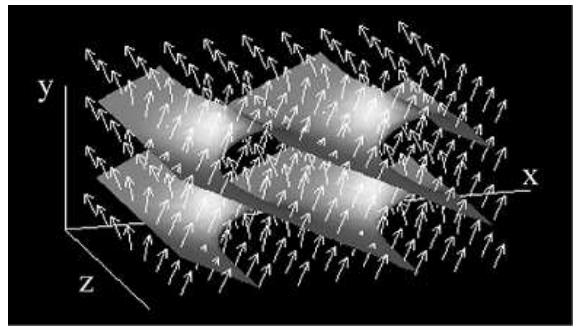}
\caption{
Snapshot of the grain boundary structure at $\tau=-0.02, g=1, B=0.2, K=0.02$
and $\alpha=50$$^\circ$. 
Plotted are the isosurface ${\rm Re}\,\Psi=0$ 
and the director $\bn$ (arrows).
}
\label{fig:SNAP}
\end{figure}
\par
We first compare the obtained layer structure with Scherk's 
first surface. The deviation from the minimal surface can be
estimated using the spatial average of $H^{2}$:
\eq
\lrA{H^{2}}=\f{1}{V}\int d\br\; H^{2}
\label{eq:H2}
\qe
This measure has more direct physical meaning 
than the previous ones~\cite{Duque} because \Eq.{H2} 
is proportional to the bending elastic energy of layers.
The mean curvature is calculated through the 
layer normal $\bem$ as $H=\div{\bem}$,
while $\bem$ is calculated through 
the phase $\Phi$ of the smectic order parameter 
$\Psi$ as $\bem = \grad{\Phi}/|\grad{\Phi}|$.
In this way we can compute the mean curvature at every point, 
which is very important for averaging out the effect of mesh size.
\begin{figure}
\includegraphics[width=83mm]{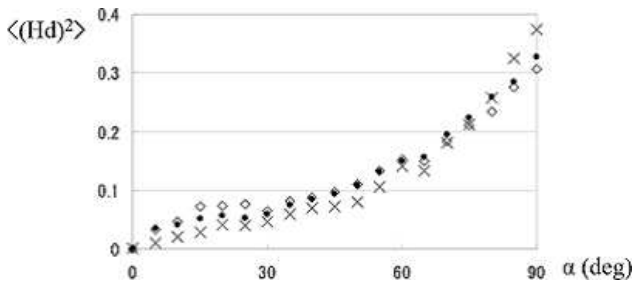}
\caption{
The spatial average of the squared and dimensionless
mean curvature $\lrA{(Hd)^{2}}$ versus
the twist angle $\alpha$
at $\tau$= ($\times$) -0.005, ($\diamondsuit$) -0.02 and ($\bullet$) -0.05.
}
\label{fig:H2}
\end{figure}
\par
The twist-angle dependence of the layer curvature 
is shown in \Fig.{H2}. 
For small $\alpha$, the TGB layer is close to Scherk's first surface,
agreeing with the analytical calculation~\cite{Kamien}.
However, $\lrA{H^2}$ grows roughly linearly as a function of $\alpha$.
For large $\alpha$, the mean curvature is not small even 
compared to the inverse layer thickness $1/d$.
\par
To understand the origin of the deviation,
we note that the layer bending elasticity 
has a contribution from Frank elasticity, through
the coupling between $\Psi$ and $\bn$. 
To see this, it would be instructive to rewrite 
the interaction term $F_{int}$ into the form
\eq
F_{int}&=&\f{B}{2}\int d\br 
\bigg| i\grad{|\Psi|} 
\nn\\
&& 
-|\Psi||\grad{\Phi}|(\bem-\bn) 
-|\Psi|(|\grad{\Phi}|-q_{0})\bn 
\bigg|^{2}.
\label{eq:Flayer}
\qe
The first term homogenizes the order parameter amplitude,
the second term locks the layer normal $\bem$ and the director $\bn$,
and the third is the layer compression term that 
adjusts the layer thickness.
Thus the coupling between $\Psi$ and $\bn$ is divided into 
the locking and layer compression terms.
If the locking effect is dominant, $\bem$ and $\bn$ will be identical.
Then the splay term in the Frank elastic energy 
is converted into the layer bending energy, as
$(K/2) \int d\br \lrS{\div{\bn}}^{2}=(K/2) \int d\br H^{2}$.
However, if the locking is weak compared to
the twist free energy ($\propto k_0$) and/or 
the layer compression energy, an unlocking of $\bem$ and $\bn$
should occur.
Then the splay term contributes less to
the effective layer-bending energy and 
$\lrA{H^2}$ can easily deviate from zero.
\begin{figure}[h]
\includegraphics[width=77mm]{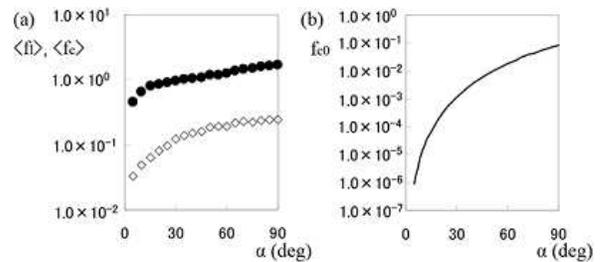}
\caption{(a) Twist-angle dependence of 
($\bullet$) the dimensionless locking energy
$f_l= (g/|\tau|) \lrM{|\Psi||\grad{\Phi}|(\bem-\bn)d}^{2}$ 
and 
($\diamondsuit$) the dimensionless layer compression energy 
$f_c=(g/|\tau|)\lrM{|\Psi_0|(|\grad{\Phi}|-q_0)d}^{2}$.
(b) The layer compression factor 
$f_{c0}=\lrM{\cos(\alpha/2)-1}^{2}$ (see text).}
\label{fig:FigCL}
\end{figure}
\par
As $\alpha$ increases, the factor $\lrS{|\grad{\Phi}|-q_0}^{2}$
in the layer compression energy greatly increases (\Fig.{FigCL}(b)),
as it is approximated as $q_0^{2} \lrM{\cos(\alpha/2)-1}^{2}$ 
at the grain boundary core. Thus the frustration between
layer compression and locking terms increases and 
leads to the unlocking of $\bem$ and $\bn$.
The spatially-averaged angle between them is over $\simeq 20$$^\circ$
for the standard parameter set and $\alpha=90$$^\circ$.
Note that the amplitude $|\Psi|$ near the grain boundary
decreases as $\alpha$ increases at higher temperature, as shown in \Fig.{PSI2}.
However, it does not affect the ratio between the two free energy 
contributions and hence is not a major cause of the deviation 
from the minimal surface. 
\begin{figure}[h]
\includegraphics[width=72mm]{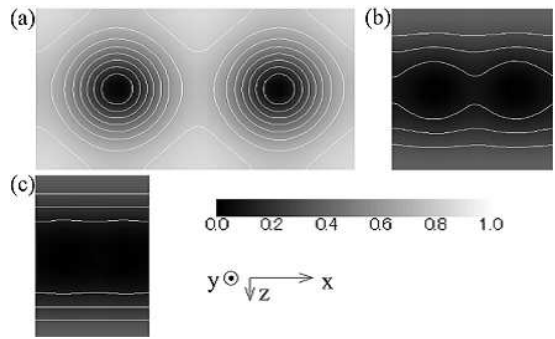}
\caption{
The squared and dimensionless order parameter $|\Psi|^{2}/(|\tau|/g)$
in the cross-section $y=\const$ at $\tau=-0.005$ and $\alpha$= (a) 30, (b) 60 and (c) 90$^\circ$.
The contour lines are drawn at 0.1, 0.2, $\ldots$, and 1.0.
For large $\alpha$, a significant melting of the smectic order
is observed even far from the dislocation cores, while it is not 
seen at lower temperature.
}
\label{fig:PSI2}
\end{figure}
\par
\begin{figure}[t]
\includegraphics[height=35mm]{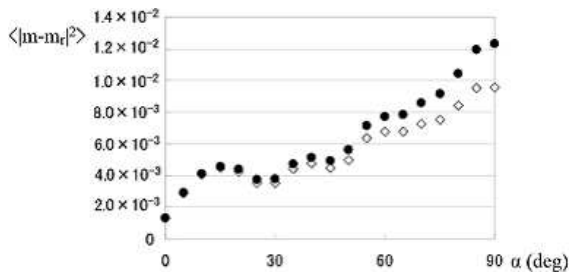}
\caption{
Deviation of the layer normal from that of
($\bullet$) Scherk's first surface and ($\diamondsuit$) LSD.
Each data set is smoothed by averaging over three consecutive points.
}
\label{fig:SL}
\end{figure}
We further tested the role of director in two ways:
(i) by decoupling the director from $\Psi$ by 
turning off the Frank elasticity,
and (ii) by adding an extra locking term $D (\bem - \bn)^2$ to the free energy.
In the unlocking limit (i), $\lrA{H^2}$ increased about five times, 
which proves Frank elasticity to be the major contribution 
to the layer bending rigidity.
With (ii), $\lrA{H^2}$ is reduced to one half for a weak extra 
locking (which reduces $\lrA{(\bem-\bn)^2}$ by only 12 \%).
These results confirm that the director unlocking plays an important role in
the deviation from the minimal surface. 
\par
Next we look at the temperature dependence.
While we varied the temperature $\tau$
(roughly proportional to $\lrA{|\Psi|^2}$) 
by a factor of ten, the deviation $\lrA{H^2}$
showed only a small change.
This also supports that the deviation is 
controlled mainly by the ratio between the
above two free energy contributions.
At high temperature, the twist free energy 
becomes important compared to the locking term, 
which is reduced due to the prefactor $|\Psi|$.
The deviation for a large twist angle is larger for 
higher temperature. This can be explained by the melting 
of the smectic order in the grain boundary, not only 
near the center of the dislocation core (\Fig.{PSI2}).
\par
Next, we compare the obtained layer structure
with both Scherk's first surface and the LSD.
To this end, we use the measure $\lrA{\lrF{\bem-\bem_{r}}^{2}}$,
where $\bem_{r} = \grad \Phi_r/|\grad \Phi_r|$ 
is the normal vector of the reference surface
defined by the phase function~\cite{NonLinear}
\eq
\Phi_r =
\tan^{\inv} \lrM{ \tan\lrS{x\sin{\f{\alpha}{2}}} \tanh{\tilde z}} + y\cos\f{\alpha}{2} - \f{\pi}{2},
\label{eq:PScherk}
\qe
where ${\tilde z}=(z\sin\alpha)/2$ for Scherk's first surface 
and ${\tilde z} = z \sin(\alpha/2)$ for the LSD.
The resultant deviations $\lrA{\lrF{\bem-\bem_{Scherk}}^{2}}$
and $\lrA{\lrF{\bem-\bem_{LSD}}^{2}}$ are plotted in \Fig.{SL}.
We see that the LSD gives a better approximation of the TGB structure
for a large twist angle,
while the difference is negligible for small $\alpha$.
\par
Finally let us compare our results with the TGB structure 
in twisted layers of block copolymer melts~\cite{Duque}.
In block copolymer melts, deviation of the intermaterial 
dividing the surface from the minimal surface is caused by 
the packing frustration~\cite{Matsen}, which corresponds 
to the layer compression energy of liquid crystals.
A self-consistent-field-theoretic calculation~\cite{Duque} 
shows that the LSD is a better model than Scherk's first 
surface, as in our case. 
However, the measure of deviation in \cite{Duque} has the dimension of 
the squared layer displacement, which causes an apparent rise of 
the deviation as $\alpha \to 0$. To compare with our results,
their measure must be multiplied by the square of the characteristic 
wavenumber proportional to $\sin^{2}(\alpha/2) \sim \alpha^2$.
Then the deviation from the LSD and minimal surfaces should 
converge to $0$ as $\alpha \to 0$.
\par
In summary, we have investigated the structure of a single 
grain boundary in the TGB phase. 
Sizable deviations from the model surfaces are obtained 
and interpreted by increase of the layer compression energy
and unlocking of the director and the layer normal.
At higher temperature and large twist angle, 
the smectic order melts even far from the dislocation core.
We hope that the core structure may be observed experimentally,
using the same kind of technique as used in the study of 
defect cores in the smectic phase~\cite{Gailhanou}.
We plan to simulate the $N_{\rm L}^*$ phase in the future, 
for which the weakening of effective layer-bending elasticity
may have some significant effect.
\par
We are grateful to Toshihiro Kawakatsu,
Helmut Brand, Jun-ichi Fukuda,
and Tomonari Dotera
for various helpful comments and discussions.
\bibliography{ogawa2.18}

\begin{thebibliography}{20}
\expandafter\ifx\csname natexlab\endcsname\relax\def\natexlab#1{#1}\fi
\expandafter\ifx\csname bibnamefont\endcsname\relax
  \def\bibnamefont#1{#1}\fi
\expandafter\ifx\csname bibfnamefont\endcsname\relax
  \def\bibfnamefont#1{#1}\fi
\expandafter\ifx\csname citenamefont\endcsname\relax
  \def\citenamefont#1{#1}\fi
\expandafter\ifx\csname url\endcsname\relax
  \def\url#1{\texttt{#1}}\fi
\expandafter\ifx\csname urlprefix\endcsname\relax\def\urlprefix{URL }\fi
\providecommand{\bibinfo}[2]{#2}
\providecommand{\eprint}[2][]{\url{#2}}

\bibitem[{\citenamefont{Renn and Lubensky}(1988)}]{Renn}
\bibinfo{author}{\bibfnamefont{S.~R.} \bibnamefont{Renn}} \bibnamefont{and}
  \bibinfo{author}{\bibfnamefont{T.~C.} \bibnamefont{Lubensky}},
  \bibinfo{journal}{Phys.\ Rev.\ A} \textbf{\bibinfo{volume}{38}},
  \bibinfo{pages}{2132} (\bibinfo{year}{1988}).

\bibitem[{\citenamefont{Goodby et~al.}(1989)\citenamefont{Goodby, Waugh, Stein,
  Chin, Pindak, and Patel}}]{Goodby}
\bibinfo{author}{\bibfnamefont{J.~W.} \bibnamefont{Goodby}},
  \bibinfo{author}{\bibfnamefont{M.~A.} \bibnamefont{Waugh}},
  \bibinfo{author}{\bibfnamefont{S.~M.} \bibnamefont{Stein}},
  \bibinfo{author}{\bibfnamefont{E.}~\bibnamefont{Chin}},
  \bibinfo{author}{\bibfnamefont{R.}~\bibnamefont{Pindak}}, \bibnamefont{and}
  \bibinfo{author}{\bibfnamefont{J.~S.} \bibnamefont{Patel}},
  \bibinfo{journal}{Nature (London)} \textbf{\bibinfo{volume}{337}},
  \bibinfo{pages}{449} (\bibinfo{year}{1989}).

\bibitem[{\citenamefont{Kamien and Lubensky}(1993)}]{NL}
\bibinfo{author}{\bibfnamefont{R.~D.} \bibnamefont{Kamien}} \bibnamefont{and}
  \bibinfo{author}{\bibfnamefont{T.~C.} \bibnamefont{Lubensky}},
  \bibinfo{journal}{J.\ Phys.\ I\ (France)} \textbf{\bibinfo{volume}{3}},
  \bibinfo{pages}{2131} (\bibinfo{year}{1993}).

\bibitem[{\citenamefont{Chan and Garland}(1995)}]{Chan}
\bibinfo{author}{\bibfnamefont{T.}~\bibnamefont{Chan}} \bibnamefont{and}
  \bibinfo{author}{\bibfnamefont{C.~W.} \bibnamefont{Garland}},
  \bibinfo{journal}{Phys.\ Rev.\ E} \textbf{\bibinfo{volume}{52}},
  \bibinfo{pages}{5000} (\bibinfo{year}{1995}).

\bibitem[{\citenamefont{Navailles et~al.}(1998)\citenamefont{Navailles, Pansu,
  Gorre-Talini, and Nguyen}}]{Navailles}
\bibinfo{author}{\bibfnamefont{L.}~\bibnamefont{Navailles}},
  \bibinfo{author}{\bibfnamefont{B.}~\bibnamefont{Pansu}},
  \bibinfo{author}{\bibfnamefont{L.}~\bibnamefont{Gorre-Talini}},
  \bibnamefont{and} \bibinfo{author}{\bibfnamefont{H.~T.}
  \bibnamefont{Nguyen}}, \bibinfo{journal}{Phys.\ Rev.\ Lett.}
  \textbf{\bibinfo{volume}{81}}, \bibinfo{pages}{4168} (\bibinfo{year}{1998}).

\bibitem[{\citenamefont{Kitzerow and Bahr}(2002)}]{Kitzerow}
\bibinfo{editor}{\bibfnamefont{H.~S.} \bibnamefont{Kitzerow}} \bibnamefont{and}
  \bibinfo{editor}{\bibfnamefont{C.}~\bibnamefont{Bahr}}, eds.,
  \emph{\bibinfo{title}{Chirality in Liquid Crystals}}
  (\bibinfo{publisher}{Springer-Verlag, New York}, \bibinfo{year}{2002}).

\bibitem[{\citenamefont{Yamamoto et~al.}(2005)\citenamefont{Yamamoto,
  Nishiyama, Inoue, and Yokoyama}}]{Yamamoto}
\bibinfo{author}{\bibfnamefont{J.}~\bibnamefont{Yamamoto}},
  \bibinfo{author}{\bibfnamefont{I.}~\bibnamefont{Nishiyama}},
  \bibinfo{author}{\bibfnamefont{M.}~\bibnamefont{Inoue}}, \bibnamefont{and}
  \bibinfo{author}{\bibfnamefont{H.}~\bibnamefont{Yokoyama}},
  \bibinfo{journal}{Nature (London)} \textbf{\bibinfo{volume}{437}},
  \bibinfo{pages}{525} (\bibinfo{year}{2005}).

\bibitem[{\citenamefont{Kamien and Lubensky}(1999)}]{Kamien}
\bibinfo{author}{\bibfnamefont{R.~D.} \bibnamefont{Kamien}} \bibnamefont{and}
  \bibinfo{author}{\bibfnamefont{T.~C.} \bibnamefont{Lubensky}},
  \bibinfo{journal}{Phys.\ Rev.\ Lett.} \textbf{\bibinfo{volume}{82}},
  \bibinfo{pages}{2892} (\bibinfo{year}{1999}).

\bibitem[{\citenamefont{Bluestein et~al.}(2001)\citenamefont{Bluestein, Kamien,
  and Lubensky}}]{Linear}
\bibinfo{author}{\bibfnamefont{I.}~\bibnamefont{Bluestein}},
  \bibinfo{author}{\bibfnamefont{R.~D.} \bibnamefont{Kamien}},
  \bibnamefont{and} \bibinfo{author}{\bibfnamefont{T.~C.}
  \bibnamefont{Lubensky}}, \bibinfo{journal}{Phys.\ Rev.\ E}
  \textbf{\bibinfo{volume}{63}}, \bibinfo{pages}{061702}
  (\bibinfo{year}{2001}).

\bibitem[{\citenamefont{Bluestein and Kamien}(2002)}]{NonLinear}
\bibinfo{author}{\bibfnamefont{I.}~\bibnamefont{Bluestein}} \bibnamefont{and}
  \bibinfo{author}{\bibfnamefont{R.~D.} \bibnamefont{Kamien}},
  \bibinfo{journal}{Europhys.\ Lett.} \textbf{\bibinfo{volume}{59}},
  \bibinfo{pages}{68} (\bibinfo{year}{2002}).

\bibitem[{\citenamefont{Santangelo and Kamien}(2006)}]{Santangelo}
\bibinfo{author}{\bibfnamefont{C.~D.} \bibnamefont{Santangelo}}
  \bibnamefont{and} \bibinfo{author}{\bibfnamefont{R.~D.}
  \bibnamefont{Kamien}}, \bibinfo{journal}{Phys.\ Rev.\ Lett.}
  \textbf{\bibinfo{volume}{96}}, \bibinfo{pages}{137801}
  (\bibinfo{year}{2006}).

\bibitem[{\citenamefont{DiDonna and Kamien}(2002)}]{DiDonna}
\bibinfo{author}{\bibfnamefont{B.~A.} \bibnamefont{DiDonna}} \bibnamefont{and}
  \bibinfo{author}{\bibfnamefont{R.~D.} \bibnamefont{Kamien}},
  \bibinfo{journal}{Phys.\ Rev.\ Lett.} \textbf{\bibinfo{volume}{89}},
  \bibinfo{pages}{215504} (\bibinfo{year}{2002}).

\bibitem[{\citenamefont{Duque and Schick}(2000)}]{Duque}
\bibinfo{author}{\bibfnamefont{D.}~\bibnamefont{Duque}} \bibnamefont{and}
  \bibinfo{author}{\bibfnamefont{M.}~\bibnamefont{Schick}},
  \bibinfo{journal}{J.\ Chem.\ Phys.} \textbf{\bibinfo{volume}{113}},
  \bibinfo{pages}{5525} (\bibinfo{year}{2000}).

\bibitem[{\citenamefont{Wit et~al.}(1997)\citenamefont{Wit, Borckmans, and
  Dewel}}]{ADEWIT}
\bibinfo{author}{\bibfnamefont{A.~D.} \bibnamefont{Wit}},
  \bibinfo{author}{\bibfnamefont{P.}~\bibnamefont{Borckmans}},
  \bibnamefont{and} \bibinfo{author}{\bibfnamefont{G.}~\bibnamefont{Dewel}},
  \bibinfo{journal}{Proc.\ Nat.\ Acad.\ Sci.\ USA}
  \textbf{\bibinfo{volume}{94}}, \bibinfo{pages}{12765} (\bibinfo{year}{1997}).

\bibitem[{\citenamefont{Memmer}(2001)}]{Memmer}
\bibinfo{author}{\bibfnamefont{R.}~\bibnamefont{Memmer}}, \bibinfo{journal}{J.\
  Chem.\ Phys.} \textbf{\bibinfo{volume}{114}}, \bibinfo{pages}{8210}
  (\bibinfo{year}{2001}).

\bibitem[{\citenamefont{Allen et~al.}(1998)\citenamefont{Allen, Warren, and
  Wilson}}]{Allen}
\bibinfo{author}{\bibfnamefont{M.~P.} \bibnamefont{Allen}},
  \bibinfo{author}{\bibfnamefont{M.~A.} \bibnamefont{Warren}},
  \bibnamefont{and} \bibinfo{author}{\bibfnamefont{M.~R.}
  \bibnamefont{Wilson}}, \bibinfo{journal}{Phys.\ Rev.\ E}
  \textbf{\bibinfo{volume}{57}}, \bibinfo{pages}{5585} (\bibinfo{year}{1998}).

\bibitem[{\citenamefont{de~Gennes}(1974)}]{deGennes}
\bibinfo{author}{\bibfnamefont{P.~G.} \bibnamefont{de~Gennes}},
  \emph{\bibinfo{title}{The Physics of Liquid Crystals}}
  (\bibinfo{publisher}{Oxford University, London}, \bibinfo{year}{1974}).

\bibitem[{\citenamefont{Kralj and Sluckin}(1993)}]{Kralj}
\bibinfo{author}{\bibfnamefont{S.}~\bibnamefont{Kralj}} \bibnamefont{and}
  \bibinfo{author}{\bibfnamefont{T.~J.} \bibnamefont{Sluckin}},
  \bibinfo{journal}{Phys.\ Rev.\ E} \textbf{\bibinfo{volume}{48}},
  \bibinfo{pages}{R3244} (\bibinfo{year}{1993}).

\bibitem[{\citenamefont{Matsen and Bates}(1996)}]{Matsen}
\bibinfo{author}{\bibfnamefont{M.~W.} \bibnamefont{Matsen}} \bibnamefont{and}
  \bibinfo{author}{\bibfnamefont{F.~S.} \bibnamefont{Bates}},
  \bibinfo{journal}{Macromolecules} \textbf{\bibinfo{volume}{29}},
  \bibinfo{pages}{7641} (\bibinfo{year}{1996}).

\bibitem[{\citenamefont{Michel et~al.}(2006)\citenamefont{Michel, Lacaze,
  Goldmann, Gailhanou, de~Boissieu, and Alba}}]{Gailhanou}
\bibinfo{author}{\bibfnamefont{J.~P.} \bibnamefont{Michel}},
  \bibinfo{author}{\bibfnamefont{E.}~\bibnamefont{Lacaze}},
  \bibinfo{author}{\bibfnamefont{M.}~\bibnamefont{Goldmann}},
  \bibinfo{author}{\bibfnamefont{M.}~\bibnamefont{Gailhanou}},
  \bibinfo{author}{\bibfnamefont{M.}~\bibnamefont{de~Boissieu}},
  \bibnamefont{and} \bibinfo{author}{\bibfnamefont{M.}~\bibnamefont{Alba}},
  \bibinfo{journal}{Phys.\ Rev.\ Lett.} \textbf{\bibinfo{volume}{96}},
  \bibinfo{pages}{027803} (\bibinfo{year}{2006}).

\end{thebibliography}
\end{document}